# *Identification of Employees Using RFID in IE-NTUA*

Rashid Ahmed
National Technical University of Athens
School of Electrical and Computer Engineering
9, Iroon Polytechniou St., 15773 Athens, Greece
e-mail: rashidis@mail.ntua.gr

John N. Avaritsiotis
National Technical University of Athens
School of Electrical and Computer Engineering
9, Iroon Polytechniou St., 15773 Athens, Greece

*Abstract*: **During the last decade with the rapid increase in indoor wireless communications, location-aware services have received a great deal of attention for commercial, public-safety, and a military application, the greatest challenge associated with indoor positioning methods is moving object data and identification. Mobility tracking and localization are multifaceted problems, which have been studied for a long time in different contexts. Many potential applications in the domain of WSNs require such capabilities. The mobility tracking needs inherent in many surveillance, security and logistic applications. This paper presents the identification of employees in National Technical University in Athens (IE-NTUA), when the employees access to a certain area of the building (enters and leaves to/from the college), Radio Frequency Identification (RFID) applied for identification by offering special badges containing RFID-tags.**

*Index Terms*- **RFID, Employees In National Technical University, Global Positioning System, Non-Line of Site.**

## I. INTRODUCTION

RECENT years have witnessed and enormous increase in moving object data from tag readings in supply chain operations, to toll and road sensor readings from vehicles on road networks. A big factor in the emergence of these data is the rapid adoption of Radio Frequency Identification (RFID) technology in industry and government. RFID is a technology that allows a sensor (RFID reader) to read from a distance and without line of sight (NLOS), a unique identifier that is provided (via a radio signal) by an "inexpensive" tag attached to an item. The technology has applications in many diverse areas, RFID offers an alternative to bar code identification that can be used in our system [1].

Generally, outdoor location estimation systems use one of two approaches: one based on the global positioning system (GPS), the other based on the cellular network system [2]. Although these approaches are convenient for positioning in outdoor environments, they do not provide accurate positioning for indoor applications because:1) the received GPS signals are too weak to provide the necessary information and, 2) the FCC Enhanced 911 [3] enables emergency services, and the localization accuracy which is within 50–300 m generally is inadequate. There is another approach for RADAR: An In-Building RF-based User location but RADAR operates by recording and processing signal strength information at multiple base stations positioned to provide overlapping coverage in the area of interest [4]. Thus, it is necessary to work other systems for indoor location-aware services to more accurately identify an indoor area.

The objective of our research is to collect the data from special badges (containing RFID tags) and identify these tags to interact with NTUA information system to detect the employees that are entering and leaving the college.

This paper is organized as follows. In Section II, we describe the background of RFID technology and problems encountered RFID systems. Section III, presents the details of RFID identification in NTUA. Section IV, we present an identification case study and the data set description of RFID-tags, finally, Conclusions are given in section V.

## II. RFID TECHNOLOGY

RFID is a technology that uses radio frequency communication to automatically identify, track and manage objects, people or animals. The devices are paired and able to "recognize" each other through the transmission of radio waves.

Implementing RFID technology will ensure the basic rights of tracking, right location, right route ,right time in and right time out, by Positive employees' identification which the standards of data exchange with confidentiality of employees' information), RFID has some desirable features, such as contactless communications, high data rate and security, NLOS readability, compactness and low cost [5], NTUA stressing guidelines for tamperproof non-transferable special badges minimizing the risk of losing transferred data. The storage technology will allow data transfer to and from host system and data storage with large storage capacity and reading ranges, RFID tags will help increase processing speed compared to bar codes. Unlike bar code, RFID-tags can be read through and around human body, clothing and non-metallic materials due to the system which utilizes radio waves provide a better approximation for location detection because of the ability of these waves to penetrate various materials. Instead of using differences in arrival times as in Ultrasound, this system utilizes signal strength to measure the location and identification [6].

### A. Radio Link Budget in RFID Systems

A typical UHF RFID system consists of RFID-reader and several passive tags .In the downlink communication, the signal transmitted on the downlink (reader to tags) contains both continuous wave (CW) and modulated commands as shown in "Fig.1". The tag responds to the reader and must demodulate the signal. The selected tags encode the data and then change the impedance of its antenna by modulating the radar cross section[7]. In the uplink communication (forward link), the reader interrogates tags with a data transfer that utilizes an ASK modulation scheme; the return data transfer, from tag to reader(addressed as downlink), utilizes a backscattered modulation scheme. In the uplink communication, the carrier signal generated by the reader is radiated out throug the antenna. The tag collects energy from the electromagnetic waves coming from the reader and converts it to DC supply for the chip. Once the tag is powered up, the reader sends the commands by modulating its carrier. After commands are completed, the reader sends an un-modulated continuous wave (CW) signal which is used to provide DC supply for the tag. More details on RFID system protocol and operation can be found [8],[9].

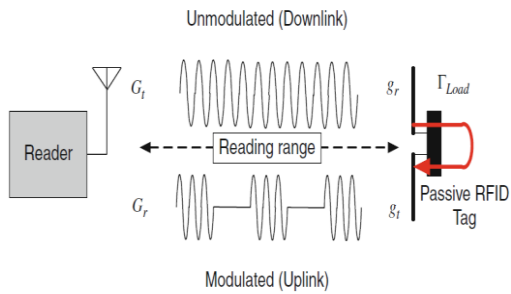

**Fig.1**.Far-field modulated backscatter

*B. Problems Encounterd in Operating RFID System*

Several potential issues must be solved successfully at the front end of the design process. Careful selection of a dynamic solution is important. In every case, a system design approach is required before implementing an RFID solution. The requirements for multiple tags, speed of operation, accuracy, cost and security must all be considered to provide the result demanded by the application[10]. Some of the common problems with generic RFID are:

*1) Reader collision:* One problem encountered with RFID systems mainly longer range UHF systems is that the signal from one reader can interfere with the signal from another where coverage overlaps. This is called reader collision. This can be avoided by using a technique named Time Division Multiple Access (TDMA) which is a special anti-collision scheme.

The readers are instructed to read at altered times, rather than both trying to read at the same time. By using this technique, RFID-reader does not interfere with each other. But by saying this, two readers that overlap each other in an area will read any RFID tag twice. Therefore the system has to be set up in such way that if one reader reads a tag, another reader does not read it again [11]. There are a lot of companies that point out how important it is that the reader collision software prevents the colliding readers from communicating with RFID tags in their respective reading zones. The Anti-Collision protocol allows the reading of large number of tagged objects at the same time and it ensures that each tag is read only once. The standard method in use is adapted by Auto-ID and it works like this; the reader asks tags to respond only if their first number of the identifier matches the number communicated by the reader. If more than one-tag responds, the reader asks for the next number in the identifier. It remains doing so until only one-tag responds. This phenomenon happens very quickly and RFID-reader can read 50 tags in less than a second. Different vendors have developed different systems for having the tags respond to the reader one at a time. Since they can be read in milliseconds, it appears that all the tags are being read simultaneously, [12].

*2)Interference:* Like other technologies using radio waves (garage door openers, remote control toys, pagers,etc.), RFID systems are subject to interference from unwanted signals electromagnetic noise. To protect against "misreads", tag data contains bits that are encoded to provide error detection by the reader to improve the reliability of the system.

*3) Presence of metal:* The presence of metal can block the performance of RFID readers and tags as well, which affects read range. However, metal can also enhance or amplify the read range of RFID with good system design.

*4) Presence of water:* The presence of water can also impede the performance of RFID, but as with electricity and water, good system design can overcome most limitations, [13].

### III. THE PROCEDURE of RFID IDENTIFICATION in NTUA

Identification of Employees can be performed utilizing RFID technology that is applied to the tracking by offering special badges which are contained RFID-tags. The tags interact with NTUA information system.

" Fig.2", shows the layout of entrances of NTUA which represents the house of RFID-reader system to read a data. Reader monitoring for enters and leaves of the employees in the building, (i.e., access to a certain area of the building), In effect, the data set is a record when the reader has read RFID-tags, and therefore, an employee's data records are the history of where and when the employees moved through the building.

RFID tags are adhesive devise which are placed on Identity Cards (ID) to be identified. When RFID-tags enters into the field of RFID antenna, they detect the activation signal, then they send certain information (based on the system setup) to the transceiver. The passive tags which are used by the RFID system do not require internal batteries and will last forever if not mistreated when employee access to a main doorway of the building "Fig.3", the data was collected from the RFID-tags which are used by the School of Computing and Information Systems at NTUA. A data set recorded when the employees have attempted to gain access to a main doorway of the building using RFID-tags and the data therefore characterizes the behavior of employees (RFID-tags) within this system.

The passive tags use the reader emissions to power a response that is usually an identification number. These tiny tags help employees reduce administration time and more importantly keep status by storing a full academic record. RFID tags can be supplied by special badges with a tamper mechanism to prevent from being removed or to emit a signal if attempted to be removed.

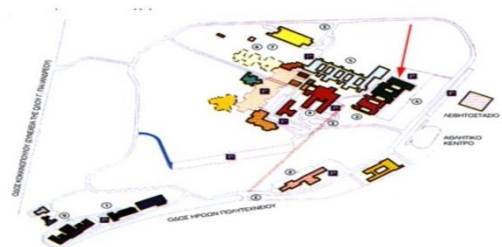

**Fig.2.** NTUA Sitemap

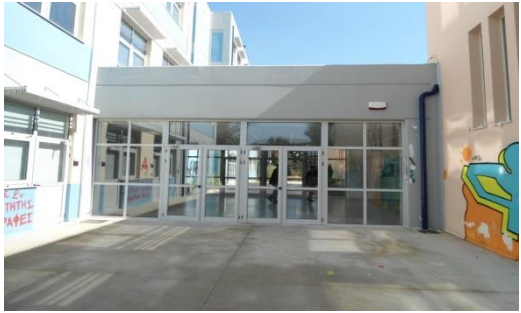

**Fig.3**. NTUA Doorway

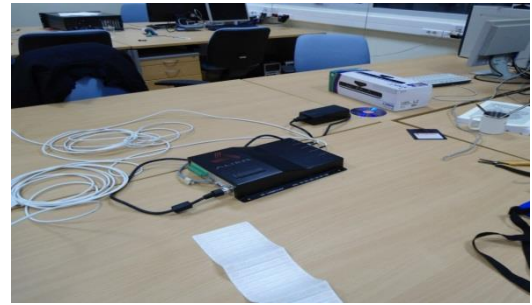

**Fig.4.** RFID system (the Alien ALR-8800 RFID reader)

The RFID-reader would be placed in specific area of a main doorway of the building in college during the employee will be located within a measurable distance. The measurable distance would be defined in the system integrator. Different materials also have an effect on the reading range and the possibility to read. Metal and water are two substances, which makes it difficult to read tags. The radio waves are absorbed by water and bounce off metal when using UHF. Low- and high frequency work better on products with water and metal than UHF and Microwave do, one drawback is that the reading range decreases when using the lower frequencies.

## IV. IDENTIFICATION CASE STUDY

The environment consists of a sensing network that helps the identification of employees/object tags within certain accuracy, and enables the wireless communication between RFID-reader and tags.

Our RFID system comprises a passive RFID-tag and the Alien ALR-8800 RFID-reader at 865.7 MHz is designed to read and program EPC and issue event reports to a host computer system. RFID-tag collects data from the transmitter signaling at frequency, and sends it to an RF receiver."Fig.4", radios are pervasive in WSNs, and adding an accurate ranging feature would enable location aware networks in ways that are not possible using other technologies [14]. The no contact and NLOS nature of this technology are significant advantages common among all types of RFID systems. All RF tags can be read despite extreme environmental factors (this study was carried out in an open field to avoid the interference of the RF noise), they can also work at remarkable speeds. In some cases, tags can be read in less than a 100 milliseconds. The other advantages are their promising transmission range and cost-effectiveness. The range that can be achieved in an RFID system is essentially determined by:

- The power available at the reader/interrogator to communicate with the tags.
- The power available within the tag to respond.
- The environmental conditions and structures (the former being more significant at higher frequencies including signal to noise ratio).

The field or wave delivered from an antenna extends into the space surrounding it and its strength diminishes with respect to distance. The antenna design will determine the shape of the field or propagation wave delivered, so that range will also be influenced by the angle subtended between tags and antenna. In space free of any obstructions or absorption mechanisms, the strength of the field reduces in inverse proportion to the square of the distance [15]. The major configuration values of software:

- Device (RF readers) setup: Used for configuring the IP addresses of the RF readers.
- Range Used for specifying what range for tags is to be scanned.
- Continuous mode: The reader will continuously report the tag ID as long as it was in the configured range.

*A. Data Set Description*

Table I shows the data records, "Fig.5", 6) show the menu research program of RFID-tags identification. Each record represents a single event; the outcome of when an employee presents a tag to the reader. RFID-reader generates one data record each time it attempts to read tags. A data record is a space-separated set of attributes and corresponding data values. i.e., data record specifies tag number 9806 as follows:

- It was granted access (enter) through the doorway by RFID-reader on date 9/2/2012 and at time 09:15:51AM.
- It was granted access (leave) at time 12:33:12 PM and it was granted access (enter) at time 01:10:11 PM.
- It was granted access (leave) at time 04:01:05 PM.

The individual fields of a data record are described in Table.I., RFID system can read the information on multiple tags simultaneously without necessarily requiring LOS and without the need for a particular orientation.

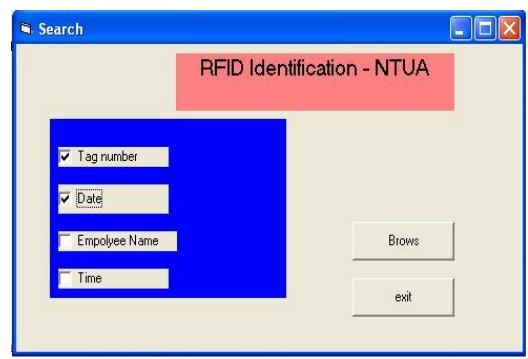

**Fig.5.** RFID tags identification program.

TABLE I
DATA RECORD ATTRIBUTES

| Tags | Time | | Access | Date |
|------|------|---|--------|------|
| 9027 | 08:33:33 | AM | pass | 9/2/2012 |
| 9030 | 09:03:10 | AM | pass | 9/2/2012 |
| 9034 | 08:50:23 | AM | pass | 9/2/2012 |
| 9806 | 09:15:51 | AM | pass | 9/2/2012 |
| 9808 | 09:10:39 | AM | pass | 9/2/2012 |
| 9030 | 11:19:44 | PM | pass | 9/2/2012 |
| 9806 | 12:33:12 | PM | pass | 9/2/2012 |
| 9806 | 01:10:11 | PM | pass | 9/2/2012 |
| 9027 | 03:30:11 | PM | pass | 9/2/2012 |
| 9030 | 03:45:45 | PM | pass | 9/2/2012 |
| 9034 | 03:55:23 | PM | pass | 9/2/2012 |
| 9806 | 04:01:05 | PM | pass | 9/2/2012 |

**Fig.6.** RFID-tags data identification.

## V. CONCLUSION

This combination of automation, identification, integration and increased accuracy has drawn attention to RFID in the employee's identification for the benefits of reduced administration time, automation of security, auditing, identification performance statistics or all of the above and reduction in any procedural errors by using a full normal record .

The data set characterizes the behavior of employees within the system as a tag acts as a surrogate for an entity. This data would be useful in analyzing the behavior of employees within this RFID system. The RFID applications of employee identification are have more impact in situations where attendance in the university needs to be monitored. The basic advantage of RFID tags over barcodes is that we can write on these tags, and automatically read many tags simultaneously even if we can't see them.

Based on the analysis of this study, future works includes applying RFID Identification /tracking for Bank customers. However, it should be taken into consideration, RF signals exhibit multipath propagation due to the environment effect from obstructing structures such as walls and corridors in the bank.

## AKNOWLLEDGMENT

This research is supported by the School of Electrical and Computer Engineering, National Technical University in Athens (NTUA), Greece.

**Rashid Ahmed** is a Ph.D student in the School of Electrical and Computer Engineering, National Technical University of Athens, Greece, and is a student member of the IEEE. E-mail: rashidis@mail.ntua.gr.

**John N. Avaritsiotis** is a Professor of Microelectronics in the Department of Electrical and Computer Engineering of the National Technical University of Athens (NTUA). He has published over 80 technical articles in various scientific journals, and has presented more than 30 papers at international conferences. His present research interests concern the development of surface micro-machining processes for the production of micromechanical sensors and design and prototyping of various types of multi-sensor systems for various applications. He is the Director of two R&D Laboratories: the Microelectronics Lab and the Electronic Sensors Lab of NTUA. He is the Co-Editor of the Journal Active and Passive Electronic Devices, Guest Editor of IEEE Transactions on Components, Packaging and manufacturing Technology, Senior Member of IEEE and Member of IOP and ISHM.